\documentstyle[epsf,11pt]{article}
\begin{document}
\begin{titlepage}
\begin{flushright}
TPR-95-29 \\
quant-ph/9511043 \\
October 1995
\end{flushright}
\vskip 0.7cm
\begin{center}
{\bf \large On Quantum Field Brownian Motion, Decoherence}
\vskip 0.15cm
{\bf \large and Semiquantum Chaos}
       \footnote{
       Supported in part by Heisenberg
       Programme (D.F.G.) and University of
       Arizona.
       }
\vskip 0.5cm
Hans-Thomas Elze
\footnote{E-mail: ELZE@CERNVM.CERN.CH}
\vskip 0.15cm
Institut f\"ur Theoretische Physik, Universit\"at Regensburg \\
93053 Regensburg, Germany. \\
Physics Department, University of Arizona, Tucson, AZ 85721, U.S.A.
\end{center}
\vskip 1.5cm
\noindent
{\bf Abstract:}
Entropy production in quantum (field) systems
requiring envi- ronment-induced decoherence
is described in a Gaussian variational approximation.
The new phenomenon of Semiquantum Chaos is reported.

\vskip 2.0cm
\begin{center}
{\it Presented at the International Conference on Nonlinear Dynamics,
Chaotic and Complex Systems, Zakopane (Poland), 7-12.11.95. \\
To be published in Acta Physica Polonica B.}
\end{center}
\end{titlepage}

\setcounter{footnote}{0}
\begin{center}
{\bf \large On Quantum Field Brownian Motion, Decoherence}
\vskip 0.10cm
{\bf \large and Semiquantum Chaos}
       \footnote{
       Supported in part by Heisenberg
       Programme (D.F.G.) and University of
       Arizona.
       }
\vskip 0.20cm
Hans-Thomas Elze
\footnote{E-mail: ELZE@CERNVM.CERN.CH}
\vskip 0.15cm
Institut f\"ur Theoretische Physik, Universit\"at Regensburg \\
93053 Regensburg, Germany. \\
Physics Department, University of Arizona, Tucson, AZ 85721, U.S.A.
\end{center}
\vskip 0.25cm
\noindent
{\bf Abstract:}
Entropy production in quantum (field) systems
requiring envi- ronment-induced decoherence
is described in a Gaussian variational approximation.
The new phenomenon of Semiquantum Chaos is reported.

\begin{center}
{\it 1. Introduction}
\end{center}

Entropy is generated in complex systems by a dynamical separation
of ``relevant'' degrees of freedom (observables) from ``environment''
modes, which we integrate out.
We generalize the Feynman-Vernon influence functional
for single-particle quantum Brownian motion to quantum field
theory \cite{I}.
Our nonperturbative approach is based on a Gaussian variational
principle, i.e. time-dependent Hartree-Fock approximation (TDHF)
in terms of Wightman functions. We calculate for the
first time the induced {\it von Neumann entropy} (vNE) in
non-equilibrium situations \cite{I2}. The functional formalism is
sufficiently general to study in field theory, e.g.,
phase transitions and environment-induced quantum decoherence effects

\cite{I,Zurek}.\footnote{For brevity we refer
to the references, from where relevant literature can be traced.}

Working on the quantum to classical transition in the long wavelength
regime of nonlinear systems, we attempt to understand
whether and to what extent classical deterministic chaos is important
for properties of quantum fields. Presently, we
report on the zero-dimensional
infrared limit of a prototype $\phi^4$ field theory, i.e. a quantum

mechanical double-well oscillator, which is regular in the classical
limit. However, including TDHF quantum corrections leads to a new
phenomenon: {\it Semiquantum Chaos} \cite{I3}. There exist
energy dependent transitions between regular behavior and
deterministic
chaos in this model. Extensions of this work may lead
towards experiments on semiquantum
chaos and the underlying quantum decoherence effect.

\begin{center}
{\it 2. Quantum Field Brownian Motion and Decoherence}
\end{center}

The long-standing ``entropy puzzle'' \cite{I}
dates back to Fermi and Landau
trying to understand the
rapid thermalization of high energy density ($\gg 1\;\mbox{GeV/fm}^3$)
matter in strong interactions. It can be
related to the concepts of {\it open quantum systems} and
{\it environment-induced quantum decoherence} \cite{I,Zurek}.
Effectively, {\it unitary time evolution} of the observed part
of the system breaks down in the transition from a quantum
mechanically pure initial state to an impure final state.
Evolution by $\exp (-i\hat{\mbox{H}}t)$ preserves the purity
of a state and, therefore, {\it cannot} produce entropy.
A complex pure-state quantum system, however, can show
quasi-classical behaviour, i.e. an impure density (sub)matrix
together with decoherence of the associated ``pointer'' states
in the observed subsystem \cite{I,Zurek}.

This has been studied
in detail in a nonrelativistic single-particle model resembling
an electron coupled to the quantized electromagnetic field,
however, with an enhanced infrared spectral density.
The Feynman-Vernon influence functional remarkably yields
in the short-time strong-coupling limit that the quantum particle
behaves like a {\it classical particle} \cite{I}:
Gaussian particle wave packets experience {\it friction}
and {\it localization}, i.e. no quantum mechanical spreading,
and their coherent {\it superpositions
decohere}. Decoherence here leads directly to vNE,
$S\equiv\mbox{Tr}\hat{\rho}\mbox{ln}\hat{\rho}$,
$\hat{\rho}\equiv particle\; dens.\; m.$ $-$
In the related third of Refs. \cite{I}
we find exponentially fast entropy production in a
quantum system which is chaotic classically.
This requires a small decohering effect, e.g.
due to vacuum fluctuations coupled in from a higher energy
scale. Contrary to previous claims, we observe
that a zero temperature environment inducing
{\it partial decoherence} is sufficient.

In field theory one naturally considers two coupled
fields instead, e.g. fast and slow modes of one
self-interacting field, respectively.
This amounts to an observable field
interacting with an environment, i.e.
{\it quantum field Brownian motion}.
In the functional Schr\"odinger picture employing
TDHF a Cornwall-Jackiw-Tomboulis type effective action and
renormalizable equations of motion are derived \cite{I}.
With the related Gaussian wave functionals the
non-equilibrium vNE for scalar fields follows,
$S(t)=-\mbox{Tr}\{\mbox{ln}(1-Y)+Y(1-Y)^{-1}\mbox{ln}Y\}$,
tracing over coordinates and $Y$ explicitly involving two-point
functions of both fields \cite{I2}.

\begin{center}
{\it 3. Semiquantum Chaos in Classically Regular Systems}
\end{center}

Following the above approach to quantum field Brownian motion,
it becomes desirable to apply and test it in simple cases. To
study the semiclassical TDHF approximation, we consider
time-dependent homogeneous configurations in a Higgs
field theory \cite{I3}. This reduces to a quantum mechanical
double-well oscillator, i.e. a closed system to begin with.

Including TDHF corrections into one-dimensional classical
equations of motion, e.g., which otherwise behave regularly
according to the Poincar\'e-Bendixson theorem, introduces
{\it additional
nonlinear terms} together with an {\it additional effectively
classical degree of freedom}, however, representing quantum
fluctuations \cite{I3}. We obtain an autonomous first-order
nonlinear flow system (four equations instead of classically
two). Trajectories in the corresponding two-dimensional
effective potential behave regularly near potential minima or
for sufficiently high excitation energy and show deterministic
chaos connected to barrier/tunneling effects
at intermediate energies. We analyze this {\it semiquantum
chaos} by Poincar\'e sections and a new frequency
correlation function related to the density matrix of the
system \cite{I3}, which both reflect the coexistence of KAM tori
and regions of irregularity. The phenomenon is not tied to this
model.

Next, we have to include
interactions with an environment.
We expect (cf. Section 2) that tuned
radiation drives the double-well oscillator,
e.g., into the semiclassical regime. Thus, we necessarily go from
a nondissipative system to an open one, in order to make semiquantum
chaos a measurable effect. Presumably,
it can be observed in experiments with irradiated solid
state quantum dots.

I thank Th. C. Blum for collaborating on the work of Section 3.

\end{document}